\documentclass[conference]{IEEEtran}
\IEEEoverridecommandlockouts

\usepackage{cite}
\usepackage{amsmath,amssymb,amsfonts}
\usepackage{algorithmic}
\usepackage{textcomp}
\usepackage{xcolor}
\usepackage{graphicx}
\usepackage{soul}
\usepackage{balance}
\usepackage[caption=false,font=footnotesize]{subfig}

\usepackage{tabularx}
\usepackage{makecell}

\usepackage[hidelinks]{hyperref}

\begin{document}
\renewcommand{\figureautorefname}{Fig.}
\newcommand{\subfigureautorefname}{Fig.}
\renewcommand{\tableautorefname}{Table}
\renewcommand{\sectionautorefname}{Section}

\title{
DG\textsuperscript{VoiC}: Speaker Clustering for Fraud Investigation under Real Call-Centre Conditions
}
\author{\IEEEauthorblockN{Muhammad Shakeel Akram}
\IEEEauthorblockA{\textit{Aston University} \\
Birmingham,UK \\
m.akram5@aston.ac.uk}
\and
\IEEEauthorblockN{Amal Htait}
\IEEEauthorblockA{\textit{Aston University} \\
Birmingham,UK \\
a.htait@aston.ac.uk}
\and
\IEEEauthorblockN{Abdul Hamid Sadka}
\IEEEauthorblockA{\textit{Aston University} \\
Birmingham,UK \\
a.sadka@aston.ac.uk}
\and
\IEEEauthorblockN{Emma Meisingseth}
\IEEEauthorblockA{\textit{Domestic \& General} \\
Wimbledon, UK \\
Emma.Meisingseth@\\domesticandgeneral.com}
\and
\IEEEauthorblockN{Karishma Jaitly}
\IEEEauthorblockA{\textit{Domestic \& General} \\
Wimbledon, UK \\
Karishma.Jaitly@\\domesticandgeneral.com}
}

\maketitle

\begin{abstract}
Insurance fraud remains costly and operationally difficult, particularly in call-centre workflows where many customer interactions begin at FNOL. While recent fraud detection methods mainly rely on structured data, text, or images, repeated speaker identity across calls remains underused as an investigative signal. This paper presents DG\textsuperscript{VoiC}, a voice clustering framework for customer verification and cross-profile speaker linking on anonymised real call-centre audio. The approach combines sensitive information-aligned anonymisation, speech-focused preprocessing, sliding-window speaker embedding extraction, and cosine similarity based clustering to identify repeated speakers under real telephony conditions. The method was evaluated on 121 recordings, with a curated reference subset of 56 samples in 22 human-agreed speaker clusters used for validation. 
The best configuration
achieved 96\% AMI, 95\% ARI, 98\% completeness, 100\% homogeneity, and 99\% V-measure. These results show that speaker clustering can provide a strong additional signal for fraud investigation by helping analysts verify speaker consistency and surface repeated voices across customers.
\end{abstract}


\vspace{-2mm}
\section{Introduction}
Insurance fraud remains a costly and persistent problem for insurers, investigators, and genuine customers \cite{govuk2024fraud,ifb2024report,naic2024fraud,aslam2022insurance,ali2022financial}. The total estimated cost of fraud in 2023 to 2024 is \pounds14.4 billion, comprising \pounds9.2 billion affecting individuals and \pounds5.2 billion affecting businesses\cite{govuk2026fraudcost}. In the UK, insurers detected more than \pounds1.16 billion of fraudulent general insurance claims in 2024, with a further 684,800 fraudulent applications prevented at policy inception \cite{abi2025fraud}. 

This creates strong demand for earlier and more reliable fraud screening, particularly in customer-facing channels.
One such channel is the call-centre, especially at FNOL. At this stage, much of the interaction takes place over the phone, which makes identity assessment difficult in real time. Frontline agents must balance customer service with basic verification, while investigators later need to review large volumes of calls to determine whether the same speaker appears across multiple customer profiles. In investigative practice, this can leave cross-case voice reuse underused.

Most existing insurance fraud systems focus on structured data, text, images, or combinations of these modalities. Recent multimodal approaches have shown gains over unimodal baselines, but they do not model repeated speaker identity across calls \cite{piehl2021classification, backlund2023detection,chang2023design,dimri2024enhancing,gangani2023ai,tarra2024ai,perumal2023innovative,banulescu2023practical,yang2023aiml,asgarian2023autofraudnet}. At the same time, research on fraudulent phone calls has largely focused on transcript content rather than speaker reuse, while real operational call-centre audio remains difficult to study because of privacy and biometric data constraints \cite{shen2024phonescam}. For example, CallCenterEN releases large-scale anonymised call-centre transcripts, but not the underlying audio, specifically because of biometric privacy concerns \cite{dao2025callcenteren}. This leaves a gap between recent advances in speaker representation learning and the practical needs of fraud investigation in insurance call workflows.


To address this gap, we introduce DG\textsuperscript{VoiC}, a voice clustering framework for anonymised real call-centre audio. The method is designed to support two related tasks: verifying whether calls linked to a customer are spoken by a consistent speaker, and identifying repeated speakers that appear across different customer profiles. Rather than acting as a standalone fraud decision system, DG\textsuperscript{VoiC} is intended to support analyst review by surfacing voice-based links under real call-centre conditions.

The main contributions of this paper are as follows:
\begin{itemize}
    \item We present DG\textsuperscript{VoiC}, a practical voice clustering approach for customer verification and cross-profile speaker linking on anonymised real insurance call-centre data.
    \item We propose a pipeline for long and variable telephony calls, combining word-level aligned anonymisation, speech-focused preprocessing, sliding-window embedding extraction, and similarity-based clustering.
    \item We evaluate the approach on a real-world dataset and show that ECAPA-based speaker embeddings can achieve strong agreement with human-reviewed speaker clusters.
\end{itemize}

The paper is organised as follows. \autoref{sec:background} reviews related work. \autoref{sec:proposedSolution} describes the proposed DG\textsuperscript{VoiC}. \autoref{sec:experimental} outlines the experimental setup, and \autoref{sec:results} presents the results and discussion. Finally, \autoref{sec:conclusion} concludes the paper.

\begin{figure*}[!t]
\vspace{-2.75mm}
    \subfloat[Non-fraudulent scenario]{%
        \includegraphics[width=0.44\linewidth]{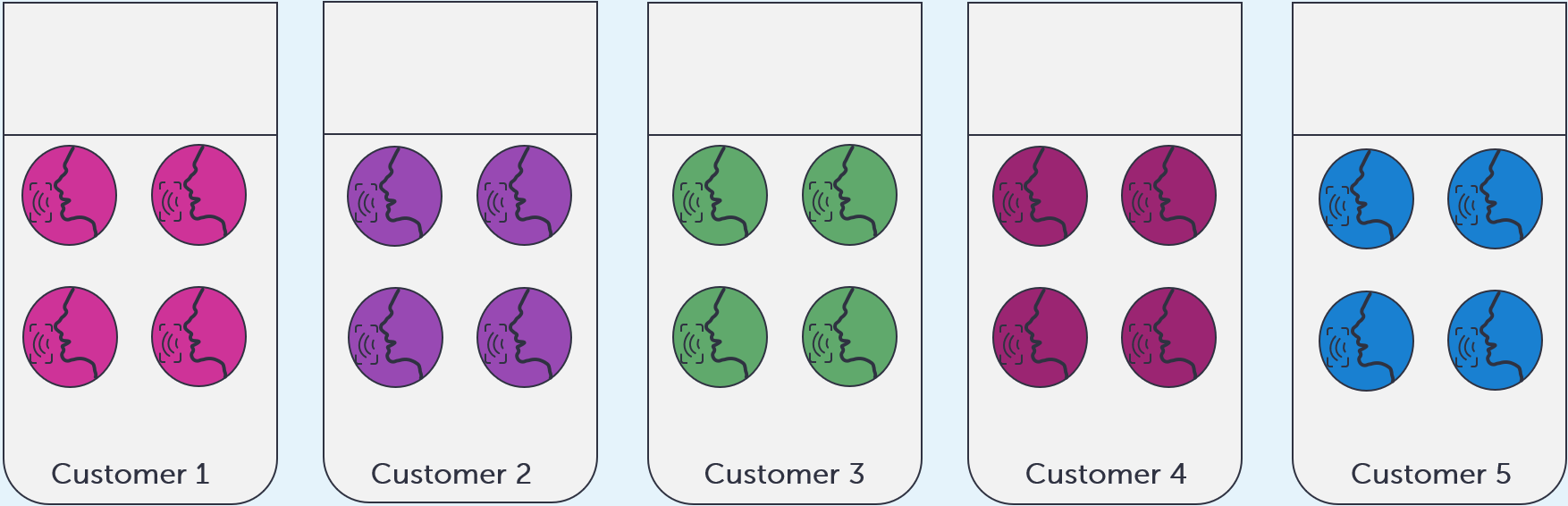}%
        \label{fig:c0_a}
    }\hfil
    \subfloat[Fraudulent scenario]{%
        \includegraphics[width=0.44\linewidth]{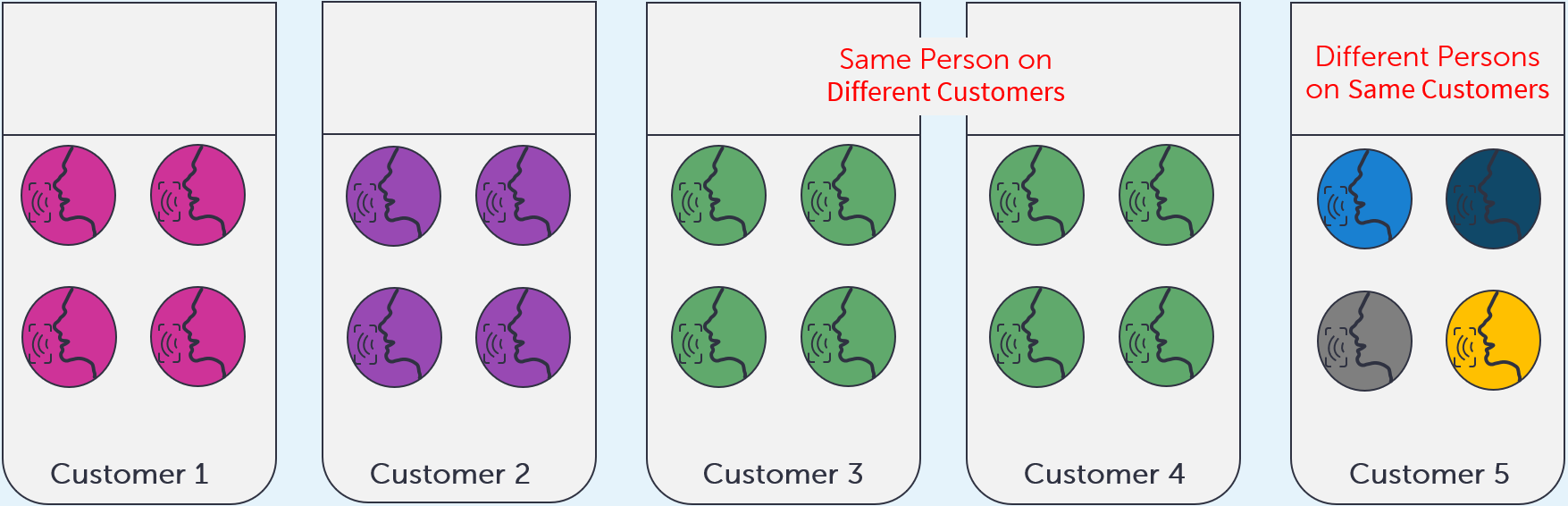}%
        \label{fig:c0_b}
    }
    \vspace{-3mm}
    \subfloat[High risk fraudulent voice prints]{%
        \includegraphics[width=0.44\linewidth]{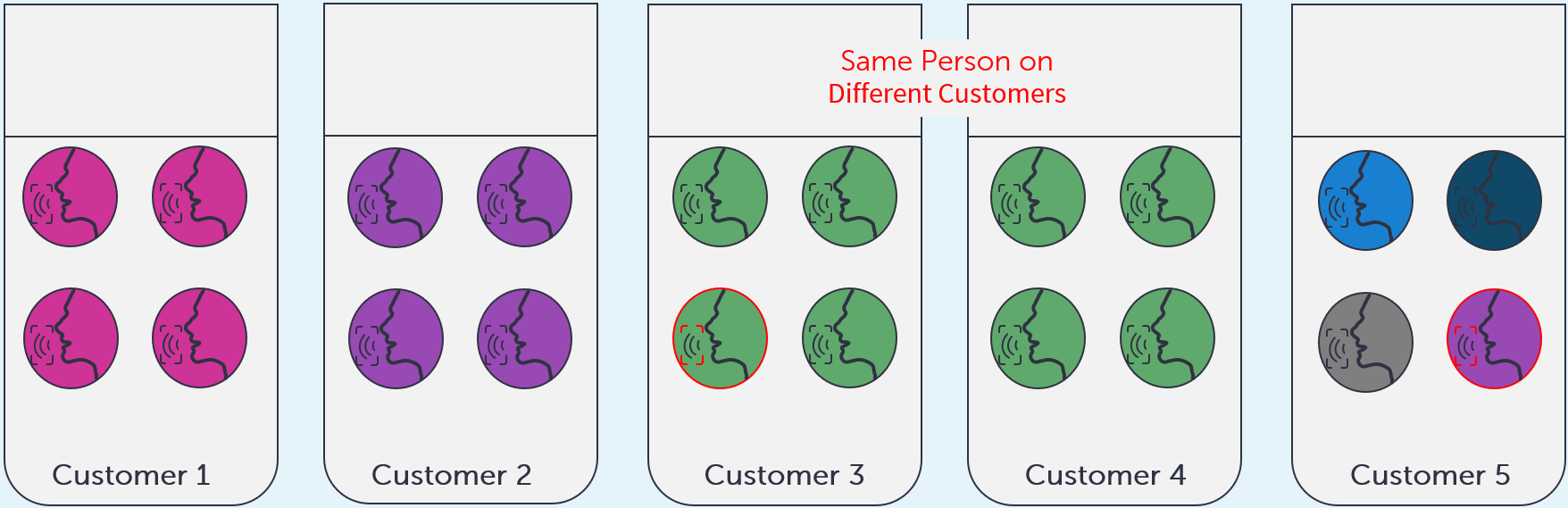}%
        \label{fig:c0_c}
    }\hfil
    \subfloat[Not selling to high-risk voice prints]{%
        \includegraphics[width=4.85cm,height=2.55cm]{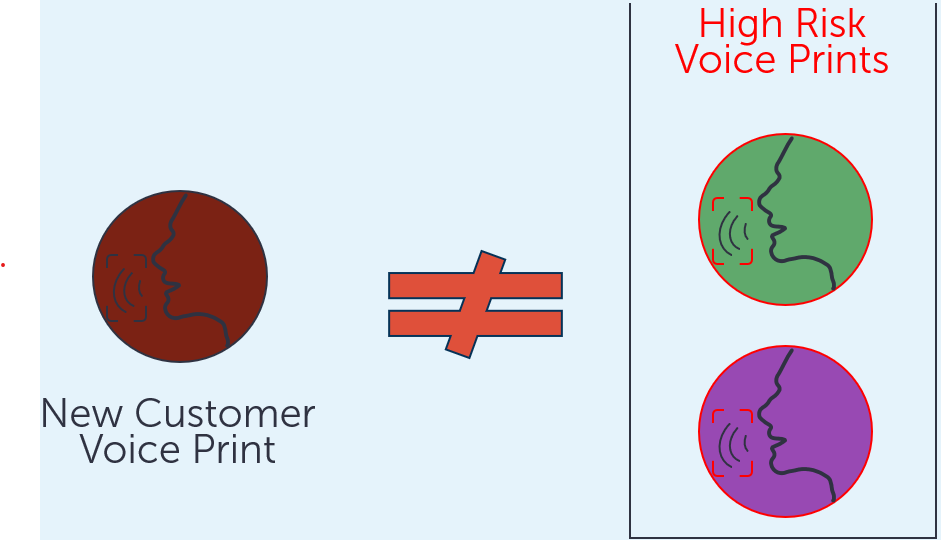}%
        \label{fig:c0_d}
    }
\vspace{-1.75mm}
    \caption{Illustrating the scenarios for customer verification and fraud identification.}
    \vspace{-7mm}
    \label{fig:scenarios}
\end{figure*}

\vspace{-2mm}
\section{Proposed DG\textsuperscript{VoiC} Voice Clustering Model}
\label{sec:proposedSolution}
The proposed DG\textsuperscript{VoiC} framework uses speaker embeddings to detect voice reuse across claims and customer profiles. The aim is to support fraud investigation under real call-centre conditions while keeping false positives low. This is important at the claims stage, where incorrect flagging can affect both customer experience and operational decision-making. In practice, the system is designed to help investigators verify whether the claimed customer is the actual speaker and to identify repeated use of the same voice across different profiles.

\autoref{fig:scenarios} illustrates the main investigation scenarios. In \autoref{fig:c0_a}, each customer profile is associated with a distinct speaker, which represents the expected non-fraud case. In \autoref{fig:c0_b}, two types of risk can appear. First, the same speaker may be linked to more than one customer profile, shown by the repeated green voice prints. Second, a customer profile may contain calls from different speakers over time. In this case, some variation is expected across genuine calls, but a clearly separate voice pattern, shown in yellow, may indicate that another person is speaking on behalf of the customer. This creates two investigation needs: verifying whether the customer is the speaker in a given call, and linking suspicious voices that reappear across different profiles.

\autoref{fig:c0_c} extends this idea to recurring high-risk voices. For example, if the voice shown in purple appears under one customer profile and later reappears under another, that voice can be treated as a repeated risk indicator. When such patterns occur with strong similarity across multiple profiles, they may point to organised misuse rather than isolated inconsistency. This supports both claim review and earlier intervention for new or existing customers.

\vspace{-2mm}
\subsection{Dataset Anonymisation}
As the study uses real call-centre data, anonymisation is applied before speaker analysis. 
PII and other sensitive attributes are detected using NER, Regex, and 
references from WhisperX word-level time stamps. The corresponding audio regions are then muted using \texttt{librosa} and \texttt{soundfile}.


\vspace{-2mm}
\subsection{Speaker Clustering}
The customer voice provides a useful signal for fraud investigation. Since operational deployment may require continuous call handling, audio is processed using an overlapping sliding-window strategy. The overlap reduces the risk of missing short but informative speech regions, while very short segments are excluded to avoid unstable embeddings.

Before segmentation, long non-speech regions are removed using Resemblyzer's \texttt{preprocess\_wav}. This reduces the effect of silence and low-information regions and keeps the speaker representation focused on active speech. Each valid segment is then encoded using ECAPA-TDNN to produce a fixed-dimensional speaker embedding. Segment-level embeddings for the same customer interaction are aggregated by mean pooling to form a final voice representation.

\begin{figure}
    \centering
    \includegraphics[width=0.65\linewidth]{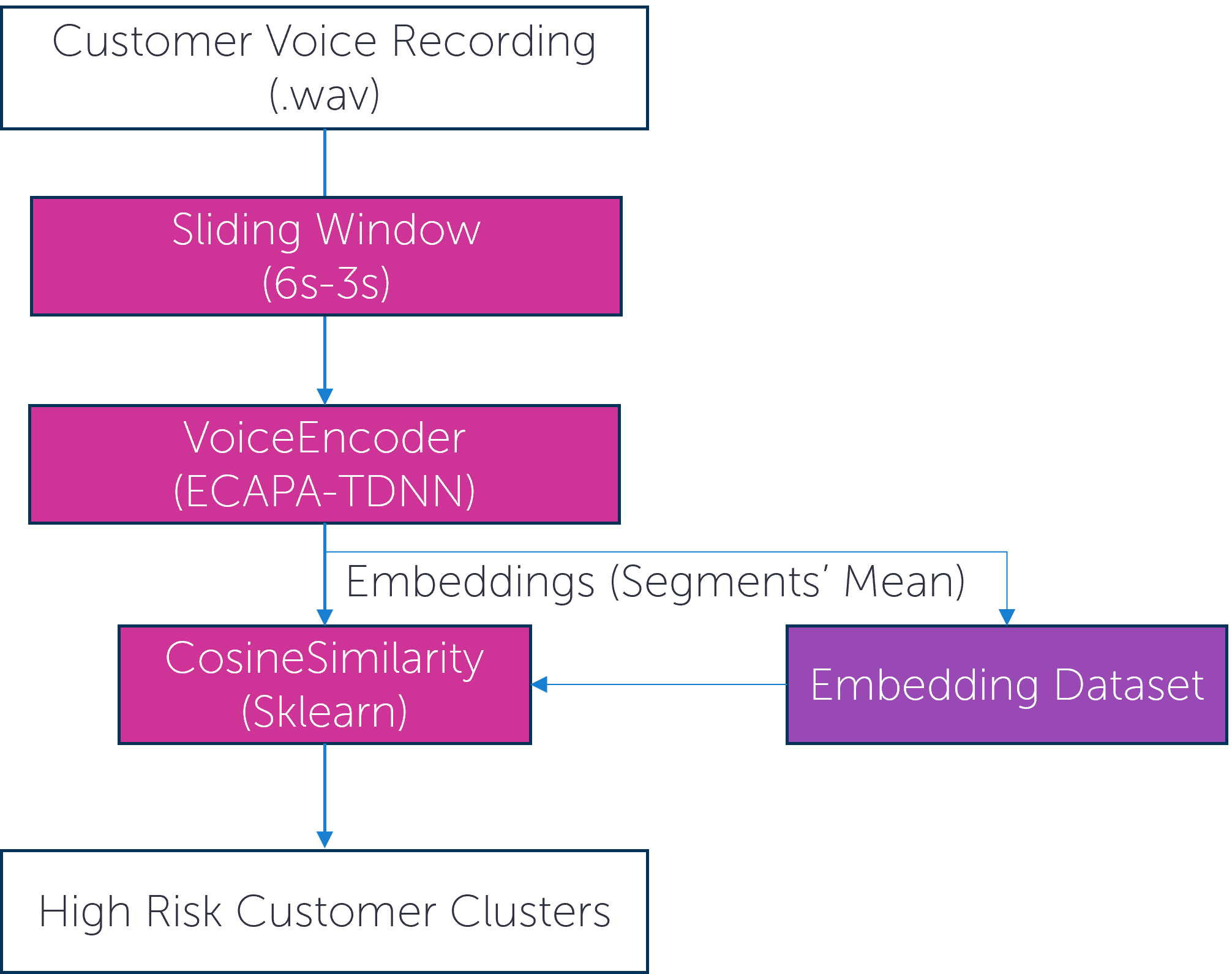}
    \vspace{-1.75mm}
    \caption{DG\textsuperscript{VoiC} model architecture.}
    \vspace{-5mm}
    \label{fig:blockdiagram}
\end{figure}

\begin{figure*}
    \centering
    \vspace{-2.75mm}
    \includegraphics[width=0.9\linewidth]{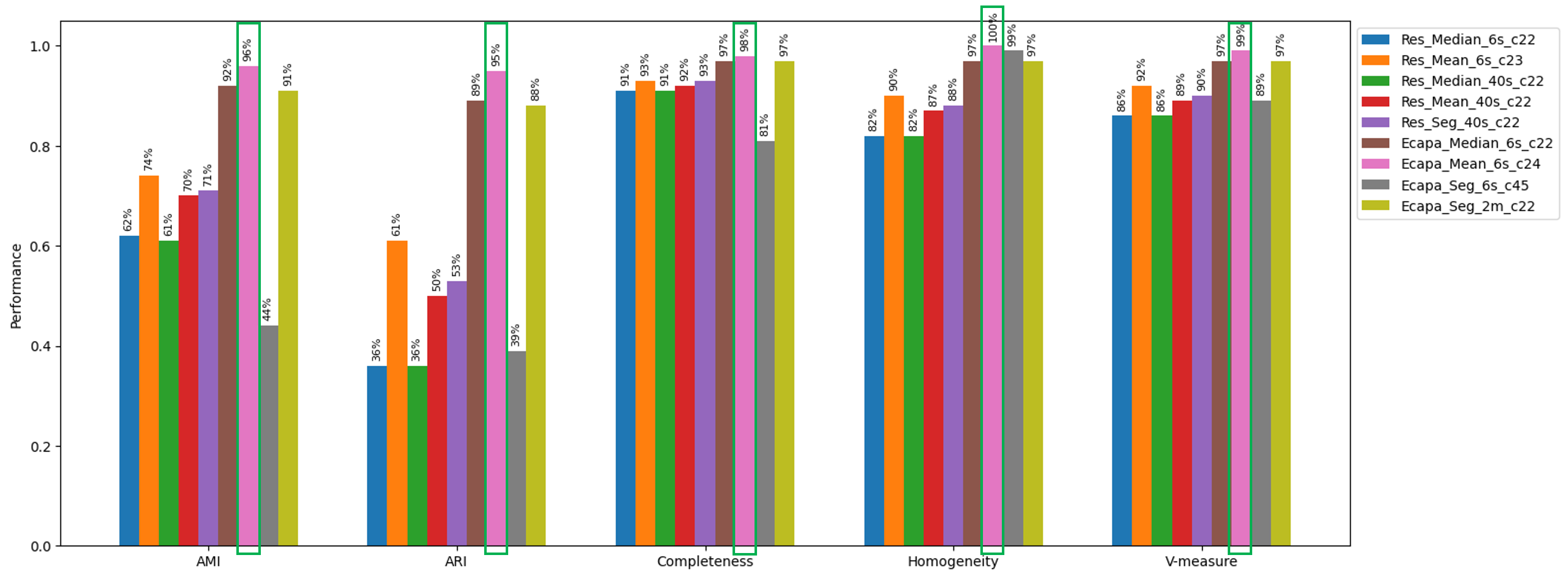}
    \vspace{-4.75mm}
    \caption{Experimentation for identifying the best combination for voice clustering model.}
    \vspace{-7mm}
    \label{fig:experiments}
\end{figure*}

Cosine similarity is then used to compare embeddings across claims and customer profiles. This makes it possible to detect repeated speakers even when the associated customer details are different. 
In fraud review, such patterns may indicate linked claims, organised misuse, or a single speaker appearing under multiple identities for human intervention and investigation. 
This detailed approach is illustrated in \autoref{fig:blockdiagram} 

To support investigator triage, speaker embeddings are further assessed using three conditions: 
\begin{enumerate}
    \item cosine similarity greater than 0.718 between candidate voice representations,
    \item membership in a cluster containing more than 4 distinct customer profiles, and
    \item cluster consistency across customer's recordings.
\end{enumerate}
The cluster-size condition is motivated by the UK Office for National Statistics report (2024), that shows over 93.6\% of households have 1-4 members \cite{ONS2025Families}. Therefore, 
legitimate voice sharing is more likely within small family units than across larger unrelated groups. 
If these conditions are met, the case ($\geq$9 profiles in a cluster and $>$92\% similarity) is assigned a higher voice-risk score to prioritise analyst review. Lower similarity, smaller cluster size, or inconsistent clustering leads to a proportionally lower score. The score is intended to prioritise analyst review rather than serve as a standalone fraud decision.

\begin{figure*}[!t]
    \centering
    \subfloat{\includegraphics[width=0.66\linewidth]{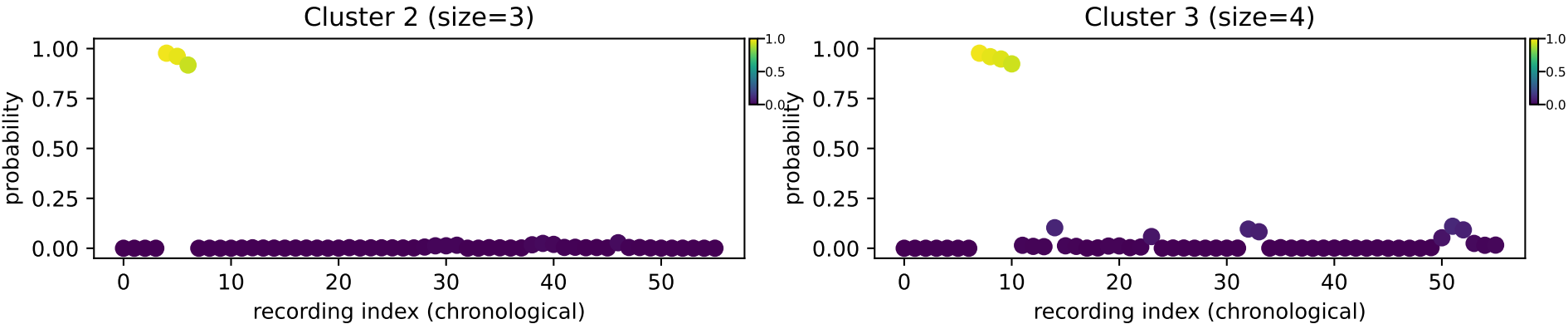}%
    \label{fig:c0}}\hfil
    \subfloat{\includegraphics[width=0.33\linewidth]{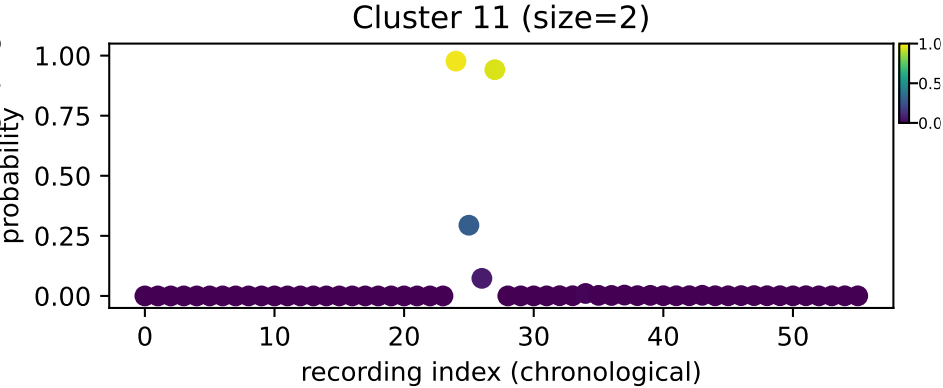}}%
    \vspace{-4mm}
    \caption{Example cluster formations on the real call-centre dataset for the best-performing configuration (Ecapa\_Mean\_6s\_c24).}
    \vspace{-5mm}
    \label{fig:clusters}
\end{figure*}

\vspace{-2mm}
\section{Experimental Setup}\label{sec:experimental}
The study used 121 real call recordings collected from multiple customer profiles. To ensure that speaker consistency could be assessed across repeated interactions, the sample was restricted to customers with at least four available recordings. This provided enough within-customer variation to evaluate where the proposed clustering approach could separate natural voice variability from true cross-profile voice reuse. 

To establish a cleaner evaluation subset, the recordings were independently blind reviewed by two fraud-domain practitioners from Domestic and General. Each reviewer listened to the calls and grouped them into speaker clusters based on perceived voice similarity. Only the samples for which both reviewers with high confidence produced the same clustering outcome were retained for validation. This resulted in a subset of 56 samples, organised into 22 agreed clusters, which was used as the clean reference set for modelling and assessment. The selected samples exhibit recording durations ranging from 2min 45s to 34min 38s, with an average duration of 10min 25s.

WhisperX \cite{bain2023whisperx} was used to generate word-level time-stamps along with a hybrid masking pipeline that combined RoBERTa-based NER \cite{roberta_large_ner_english} with rule-based Regex to detect personally identifiable and sensitive information. The corresponding audio spans were muted using \texttt{librosa} and \texttt{soundfile}, producing an anonymised version of each call for downstream analysis.
For speaker modelling, ECAPA-TDNN \cite{Ecapa} was used to extract embeddings. Cosine similarity using scikit-learn \cite{scikit-learn}, was then used to compare embeddings and identify potential voice reuse across customers.


Latency evaluation was performed using \texttt{\%\%timeit} with 20 repeated runs. Experiments were executed on Databricks custom compute (runtime 17.3 LTS ML) with 32 CPU cores, 128\,GB memory, and a single 64\,GB GPU for both driver and worker nodes.


\vspace{-2mm}
\section{Results and Discussion}\label{sec:results}
This section reports the clustering performance of the proposed DG\textsuperscript{VoiC} pipeline on the curated evaluation subset. The aim was to identify the configuration that best separates natural within-speaker variation from true cross-profile voice reuse under real call-centre conditions.

\autoref{fig:experiments} summarises the best result from each experiment. The evaluation compared two speaker embedding models, Resemblyzer VoiceEncoder (Res) and ECAPA-TDNN (Ecapa), under several segmentation settings. These included overlapping sliding windows of 6\,s with 3\,s hop size and 40\,s with 20\,s hop size, as well as single-segment processing using 6\,s, 40\,s, and 2\,min portion. Segment-level embeddings were then aggregated using either mean or median pooling. Clustering was primarily based on cosine similarity with the clustering threshold varied between 0 and 1 in increments of 0.001. DBSCAN and FAISS were also explored, but both performed worse on this real dataset and were therefore excluded from the final model comparison.

Across all tested settings, the results (\autoref{fig:experiments}) were generally strong, which suggests that speaker embeddings remain informative even under noisy and operationally variable call-centre conditions. The best overall configuration used Resemblyzer preprocessing for silence removal, ECAPA-TDNN for embedding extraction, a 6\,s sliding window, mean pooling, and a cosine similarity threshold of 0.718 (selected on the experimental sweep as the best development threshold). This setup achieved 96\% adjusted mutual information (AMI), 95\% adjusted Rand index (ARI), 98\% completeness, 100\% homogeneity, and 99\% V-measure. Leading to 95\% accuracy and 0.96 F1 score. 

\begin{table*}[t]
    \caption{Contextual comparison with technically closest approaches.} \vspace{-2.75mm}
\begin{tabular}
{|@{\hspace{0em}}p{0.04\textwidth}|@{\hspace{0.3em}}p{0.2\textwidth}|@{\hspace{0.3em}}p{0.25\textwidth}|@{\hspace{0.3em}}p{0.45\textwidth}|}
\hline
\textbf{Ref}&\textbf{Task Focus}&\textbf{Performance}&\textbf{Comments}\\
         \hline
         \textbf{DG\textsuperscript{VoiC}}&Customer-level speaker clustering across
calls for fraud investigation&AMI:96\%,  ARI:95\%, C:98\%, H:100\%, V:99\%, A:95\%, F1: 0.96, L: 10.08s, EER: 3.85\%, FAR: 0.5\%,  FRR:9.62\%&Low–medium complexity; Resemblyzer-ECAPA-Cosine
based pipeline designed using real data for real call-centre workflows\\
         \hline
         \cite{sonwane2024trustcaller}&Caller identification and authentication&A: 89\%, F1:0.88&MFCC-SVN\_linear based supervised models;
medium complexity;  6-7s clips\\
         \hline
         \cite{kim2025multimodal}&Multimodal voice phishing and synthetic
voice detection&[Voice-only] A: 73\%, *L:32.4s&CNN-BiLSTM with MFCC features;
medium–high complexity; 20–170s synthetic audios\\
         \hline
        \cite{brydinskyi2024comparison}&Speaker verification&EER:1.7\%, FAR:1.7\%, FRR:1.7\%, DCF:0.1\%& ECAPA-based solution on  non-english Synthetic data\\
         \hline
         \cite{nandal2025secureasv}&Speaker verification and clone detection &Random Forest EER:4.4\%, \quad \quad Neural Network A:96\%&Dual-model pipeline; higher complexity;
ASVspoof data\\ 
         \hline
         \cite{Sholokhov2023}&Online speaker recognition and clustering&EER: 4.98\%, DER: 3.32\%&sph-PLDA-based probabilistic back-end; medium complexity;\\
         \hline
        \cite{alvarez2025feature}&Speaker diarization (telephone speech: CALLHOME)&[2-4 Speakers] DER: 7.2\%, A: 68.8\%& Oracle VAD, ECAPA-TDNN, Mel-filterbank features, EEND-EDA based Neural diarization; higher modelling
and deployment complexity\\
        \hline
        
                  \multicolumn{4}{l}{\footnotesize \makecell{DCF: Detection Cost Function, AHC: Agglomerative Hierarchical Clustering.
                  A: Accuracy, C: Completencess, H:Homogeneity, V:V-measure, \\L: Latency/recording
                  *Compute: Intel i9-9900K CPU-RTX 3090 GPU-24 GB memory.}}\\
\end{tabular}
\label{tab:comparison}
\vspace{-5mm}
\end{table*}

The best-performing configuration produced 24 clusters, very close to 22 clusters in the labels. A closer review showed that the additional clusters were not random errors. In one case, a customer with four recordings had two calls that sounded noticeably different from the other two on re-listening, which led to a split cluster. In another case, one recording contained sustained loud background noise in the later part of the call, which shifted its representation away from the remaining calls from the same customer. 
This highlights residual acoustic variability as an important target for future work.

For post-hoc operational scoring, cosine similarity was also mapped to a bounded probability score using a sigmoid function 
to support ranking and triage. 
\autoref{fig:clusters} shows example cluster formations from the best configuration. Overall, the results indicate that the proposed configuration can group repeated speakers with high consistency and can support fraud investigation by surfacing cross-profile voice links for analyst review.


In this setup, embedding extraction required an average of 10s/recording (9min 26s for the full dataset), while clustering and customer linking took 84ms/recording (4.69s in total). This results in an average end-to-end processing time of 10.08s/recording.


As most technically related studies report verification- or diarization-oriented metrics such as Equal Error Rate (EER). Although the primary task addressed here is customer-level speaker clustering across calls hence clustering agreement metrics remain the primary evaluation criterion. We additionally derive an auxiliary verification-style analysis for contextual comparison only. This yields 3.85\% EER, with a false acceptance rate (FAR) of 0.50\% and 9.62\% false rejection rate (FRR) at 0.718 clustering threshold.


\vspace{-2mm}
\section{Related Work} \label{sec:background}

Existing work on fraudulent phone calls is closest in operational setting but primarily focuses on transcript content, dialogue semantics, or scam classification rather than persistent speaker identity. In parallel, most insurance fraud research relies on structured or multimodal claim data, while speech-focused studies typically address speaker diarization or verification on benchmark corpora rather than real call-centre fraud workflows \cite{yang2023aiml,asgarian2023autofraudnet,shen2024phonescam,serafini2023experimental,Sholokhov2023}. Recent work on scam automation further shows that modern speech synthesis and recognition technologies can be misused to scale impersonation attacks, reinforcing the need for voice-based controls in telephony environments \cite{gressel2024llmscam}.

Within call-centre contexts, the closest application-level studies address customer identification and caller authentication using voice biometrics. These approaches typically rely on MFCC-based features or supervised speaker identification models and aim to verify known callers rather than to cluster repeated speakers across customer profiles for fraud investigation \cite{khan2024identification,sonwane2024trustcaller,kim2025multimodal}. As a result, their evaluation protocols, reported metrics, and deployment objectives differ substantially from customer-level speaker clustering.

To the best of our knowledge, no prior work has reported the same end task addressed by DG\textsuperscript{VoiC}, namely customer-level voice clustering across calls under real insurance call-centre conditions to support speaker consistency checking and cross-profile linking. Consequently, direct numerical comparison with the state of the art is not methodologically appropriate, as published systems are evaluated on different tasks, datasets, and metrics, such as DER or EER, rather than clustering agreement measures.


For contextual comparison, Table~\ref{tab:comparison} summarises the closest technical paradigms in terms of task alignment, reported performance, and relative computational complexity. While end-to-end diarization and verification systems often achieve strong benchmark results, they typically incur higher modelling and deployment complexity. In contrast, DG\textsuperscript{VoiC} adopts a simpler embedding-based pipeline designed to operate efficiently under real-world call-centre constraints and to support analyst-led fraud investigation rather than automated decision-making.

\vspace{-2mm}
\section{Conclusion}\label{sec:conclusion}

This paper presented DG\textsuperscript{VoiC} for customer verification and cross-profile speaker linking in insurance fraud investigation. Experiments on anonymised real call-centre audio showed that speaker embeddings can consistently group repeated speakers under practical telephony conditions, providing a useful signal for analyst-led fraud review.
Future work will focus on improving robustness to challenging call conditions, including background noise, within-speaker variability, and customer-side speaker diversity. Additional extensions will explore complementary audio cues, such as prosodic and behavioural features, to enhance voice-based risk scoring alongside speaker clustering, with validation guided by expert assessment against real-world fraud outcomes.



\balance
\bibliographystyle{IEEEtran}
\bibliography{Ref}

\end{document}